# Accelerated Electromagnetic Simulation of MRI RF Interactions with Graphene Microtransistor based Neural Probes for Electrophysiology–fMRI Integration

Suchit Kumar[1], Alejandro Labastida-Ramírez[2], Samuel M. Flaherty[2], Anton Guimerà Brunet[3, 4], Nerea Alvarez de Eulate[3, 4], Kostas Kostarelos[2], Ben Dickie[5], Rob C. Wykes[1,2], Louis Lemieux[1,*]

[1] University College London, Queen Square Institute of Neurology, London, United Kingdom

[2] Centre for Nanotechnology in Medicine & Division Neuroscience, University of Manchester, Manchester, United Kingdom

[3] Instituto de Microelectrónica de Barcelona, IMB-CNM (CSIC), Cerdanyola del Vallès, Spain

[4] Centro de Investigación Biomédica en Red en Bioingeniería, Biomateriales y Nanomedicina (CIBER-BBN), Madrid, Spain

[5] Division of Informatics, Imaging and Data Sciences, University of Manchester, Manchester, UK

[*]Author to whom any correspondence should be addressed.

E-mail: l.lemieux@ucl.ac.uk

*Abstract*— *Objective*: Implementing electrophysiological recordings within an MRI environment is challenging due to complex interactions between recording probes and MRI-generated fields, which can affect both safety and data quality. This study aims to develop and evaluate a hybrid electromagnetic (EM) simulation framework for efficient and accurate assessment of such interactions. *Methods:* A hybrid EM strategy integrating the Huygens' Box (HB) method with sub-gridding was implemented FDTD solver (Sim4Life). RF coil models for mouse and rat head were simulated with and without intracortical (IC) and epicortical (EC) graphene-based micro-transistor arrays. Three-dimensional multi-layered probe models were reconstructed from two-dimensional layouts, and transmit field ($B_1^+$), electric field (E), and specific absorption rate (SAR) distributions were evaluated. Performance was benchmarked against conventional full-wave multi-port (MP) simulations using Bland–Altman analysis and voxel-wise percentage differences. *Results:* HB simulations reduced computational time by approximately 70–80%, while preserving spatial patterns of $|B_1^+|$, $|E|$, and SAR, including transmit-field symmetry and localized high-field regions. Deviations from MP were minimal for $|B_1^+|$ (median Δ% 0.02–0.07% in mice, −3.7% to −1.7% in rats) and modest for $|E|$ and SAR, with absolute SAR values remaining well below human safety limits. Graphene-based arrays produced negligible effects on RF transmission and SAR deposition. *Conclusion:* The HB approach enables computationally efficient, high-resolution evaluation of EM interactions involving microscopic probes in MRI environments, supporting simulations that are otherwise impractical with full-wave MP modeling. *Significance:* HB provides a reliable framework for preclinical safety assessment and design optimization of graphene-based implants, facilitating accelerated translation of microscopic implant simulations toward human-scale applications in the future.

*Keywords:* Simultaneous electrophysiology-fMRI, Huygens' Box method, EM simulation, MRI safety, graphene-based neural arrays, RF circuit co-simulation.

## 1. Introduction

Simultaneous electrophysiological and functional MRI in preclinical research provides complementary insights into brain activity and connectivity due to its complementary advantages, leading to higher spatiotemporal resolution (Ritter and Villringer 2006) that enables the precise mapping of haemodynamic changes related to neuronal activity, enabling the characterisation of the neurophysiological basis of fMRI signals, including the blood oxygen level-dependent (BOLD) effect and neurovascular coupling (Parker *et al* 2002, Logothetis *et al* 1999, Pan *et al* 2010). Several applications of such have been reported in animal models to study various brain diseases like epilepsy, stroke (Pirttimäki *et al* 2016, Weber *et al* 2008); irritative zone modelling (Deshmukh *et al* 2018); and pharmacological effect on neural activity (Jaime *et al* 2018). The main challenges with the animal sized MR-compatible electrophysiology systems are firstly that they are constrained by the dimension and material of the electronics, and secondly limited by the number of channels that can be placed on the head (Sumiyoshi *et al* 2011). While high-density arrays are preferred for high spatial resolution recordings, their insertion can cause large displacement leading to brain tissue injury (Zhang *et al* 2016).

Technical challenges associated with simultaneous electrophysiology-fMRI systems arise from electromagnetic (EM) and mechanical interactions between electrophysiology components and the MRI scanner, leading to subject safety concerns and degradation in data quality due to artifacts (Lemieux *et al* 1997, Hawsawi *et al* 2017, Allen *et al* 2000, Ihalainen

*et al* 2015). As these electrodes or probes are usually metallic (e.g., gold, silver, platinum, iridium, titanium) they can cause significant image distortions and signal loss in MRI images, making brain tissue visualization difficult (Stevens *et al* 2007). Also, they can cause RF-induced heating in the tissue surrounding metallic electrodes due to an increase in induced current during MRI acquisition, which in turn leads to a rise in temperature, posing potential safety risks (Erhardt *et al* 2018, Bassen *et al* 2006).

Recently, a novel electrophysiological recording technique based on graphene technology has been developed, known as Graphene Solution-Gated Field-Effect Transistor (gSGFET) arrays (Hébert *et al* 2018, Masvidal-Codina *et al* 2019). These devices feature a graphene sensing area that acts as a channel, connecting to two metal terminals (source and drain). The graphene channel directly interfaces with the biological tissue, which changes its conductivity based on the surrounding neural electrical activity. Such transistor-based configuration offers several advantages over existing technology (Bonaccini Calia *et al* 2022). These advantages include high-fidelity, wide-bandwidth brain signal detection capabilities, ranging from direct current (DC) to high-frequency activity, and the potential for multiplexing technology (Jafarian and Wykes 2022, Cisneros-Fernández *et al* 2021, Schaefer *et al* 2020, Garcia-Cortadella *et al* 2020). Furthermore, they are in sub-microscopic dimensions and contain a very small amount of metal. The recording of brain activity using such arrays (Wykes *et al* 2022) combined with concurrent fMRI acquisitions could greatly advance the understanding of brain pathology in diverse neurological disorders and injuries. However, this necessitates the investigation of the MR compatibility and safety of gSGFET arrays in the MRI environment.

Computational modeling and simulations are often performed to assess safety and the optimization of the RF hardware design. Studies have already been reported on such simulations to investigate the interaction of implants in the MRI environment by evaluating the effect on EM fields (i.e., $B_1$- and E-fields), deposited RF energy (calculated as the specific absorption rate, SAR[1]) in digital animal models (Ho 2001, Angelone *et al* 2004, Kainz *et al* 2006a, Gosselin *et al* 2014). EM interactions between the RF coil, specimen being imaged and implant can be simulated by solving Maxwell's equations using computational techniques such as the finite element method (FEM), finite-difference time-domain (FDTD) method, and method of moments (MoM), as well as hybrid approaches such as FEM/FDTD or MoM/FDTD (Guclu *et al* 1997, Ibrahim *et al* 2000, Jyh-Horng Chen *et al* 1999, Chen *et al* 1998). FDTD is the most versatile and widely used solver due to its ability to cover a wide range of frequencies, incorporate nonlinear material properties, and simulate the whole configuration simultaneously which includes the RF source, tissue models, and implant (Taflove *et al* 2005, Liebig *et al* 2013, Kabil *et al* 2016). However, simulations involving implants can be particularly computationally demanding if their size is small relative to the RF coil and specimen being imaged. Since FDTD employs a global grid-based mesh, a small implant necessitates a very high-resolution mesh, significantly increasing computational time and special hardware requirements such as GPUs (Cabot *et al* 2013). This difficulty is amplified for small implants consisting of multiple sub-units, such as gSGFET arrays, in which each sub-unit is in the order of few hundreds of microns (Bonaccini Calia *et al* 2022), while the rat's brain dimensions range from 16–20 mm (short axis) to 23–25 mm (long axis) (Paxinos and Watson 2004).

---

[1] SAR is a measure of the RF power absorbed by a given mass of tissue when exposed to an RF field and is directly proportional to the square of the electric field, expressed in watts per kilogram (W/kg) and is the basis for the main safety criterion in MRI in the absence of implants.

Ensuring an accurate SAR distribution is crucial for complying with RF safety standard guidelines (IEC 60601-2-33 2022). For efficient modeling and simulation of the EM fields and SAR around small implants, an FDTD-based Huygens' approach was shown to enable a sufficiently high-resolution mesh while assuming negligible backscattering from the electrodes to the source (Neufeld *et al* 2009). However, a major limitation of applying the Huygens' approach directly to microscopic (sub-micrometric) implants is the extremely high computational demand, often exceeding >1000 hours on a single GPU even with low grid resolution, making it impractical. Non-uniform gridding techniques, such as sub-cell and sub-gridding methods, designed to enhance the computational efficiency of the Huygens' approach and improve resolution have been explored in antenna characterization which includes perfect plane wave injection, and Ground Penetrating Radar (GPR) antenna modelling in the wide band at very high frequency (Stuchly *et al* 2000, Huang *et al* 2013, Hartley *et al* 2018). Sub-cell methods modify the coefficients in the structured grid based on integration paths corresponding to sub-cell geometries, but they are band-limited and narrow-band, restricting their generalization (Hartley *et al* 2018). Sub-gridding methods offer advantages such as the ability to divide the computational domain into different grid sizes, making them more flexible for complex modeling scenarios (Hartley *et al* 2018).

Therefore, in this work, to address the challenge of impractical simulation times and sub-optimal accuracy in systems comprising probe/arrays elements many orders of magnitude smaller than the RF coil, an accelerated hybrid EM simulation methodology is implemented and evaluated in the context of the MRI compatibility assessment of micrometre-scale Ultra-thin graphene-based probes.

## 2. Methods

The approach presented in this work integrates the Huygens' Box (HB) method with the sub-gridding technique for EM simulations for the first time using an FDTD solver on Sim4Life (ZMT 2025). The performance and accuracy of the presented method in estimating EM fields and SAR are evaluated by comparing its results with those obtained from full-wave multi-port (MP) FDTD simulations.

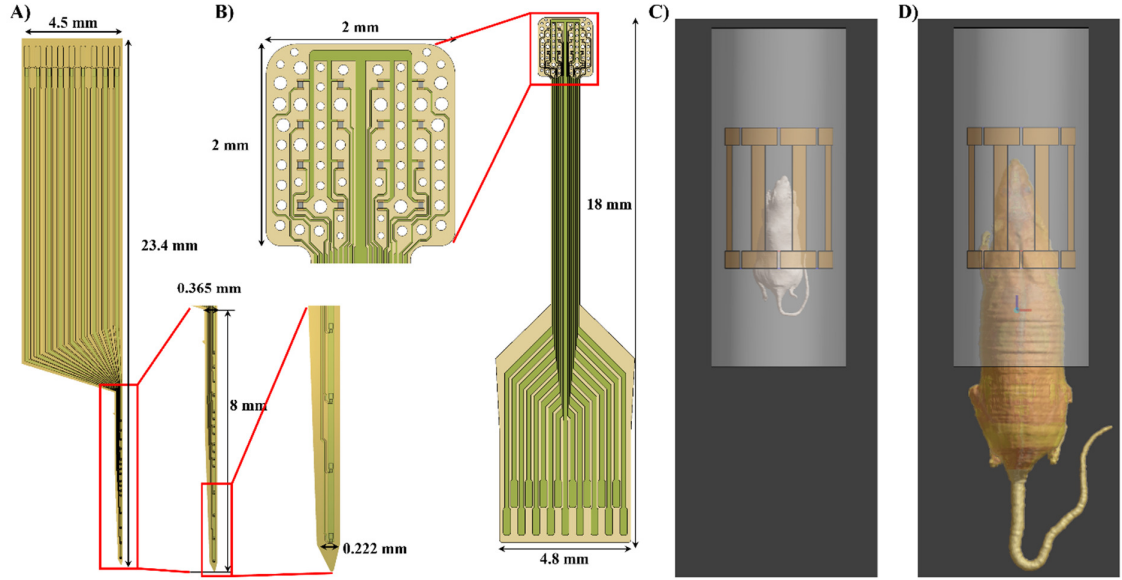

Figure 1. Schematic diagrams and dimensions of the: (A) IC; (B) EC. A detailed description on the schematic and operation of the gSGFET arrays are described in the previous works (Masvidal-Codina *et al* 2019, Bonaccini Calia *et al* 2022). Due to Sim4Life limitations in simulating graphene's atomic monolayer thickness, a minimum thickness of 100 nm was implemented in this work. (C) and (D) show the mouse and rat model, respectively.

*2.1 gSGFET Array Models*

The graphene-based arrays are characterized by their multi-channel configuration, compact design, and multilayered flexible structure with micrometer thickness (Figure 1). Two 16-channel graphene micro-transistor array designs were simulated: an intracortical (IC) type inserted within the brain tissue, and an epicortical (EC) type positioned on the cortical surface, adapted from the previous work and are being fabricated at the clean room facilities of IMB-CNM (Masvidal-Codina *et al* 2019). The IC array comprised seven layers arranged from bottom to top as follows: polyimide structural layer (L1), first metal (L2; Ti/Ni/Au), polyimide (L3; L1 minus VIAs), second metal (L4; Ti/Au), graphene monolayer patch (L5), third metal (L6; Ni/Au), and SU8 passivation (L7). The overall dimensions were 23.4 mm in length, 4.5 mm in width, and 0.01196 mm (11.96 µm) in total thickness. Similarly, the EC array consisted of five layers: polyimide structure (L1), first metal (L2; Ti/Ni/Au), graphene monolayer patch (L3), second metal (L4; Ni/Au), and SU8 passivation (L5), with overall dimensions of 18 mm in length, 4.8 mm in width, and 0.00984 mm (9.84 µm) in total thickness.

For EM modelling (Figure 2), 3D multi-layered CAD models were generated from 2D drawings of the arrays as follows: First, using *Rhino* (V8, Washington, DC, USA), the individual layers were separated and exported in DWG format. Second, these layers were imported and converted into 2D surfaces using *Sim4Life*. Finally, the 2D surfaces are thickened and combined in Sim4Life to construct a multilayered 3D model. The electrical property of graphene is based on the Drude model of graphene (Hanson 2008).

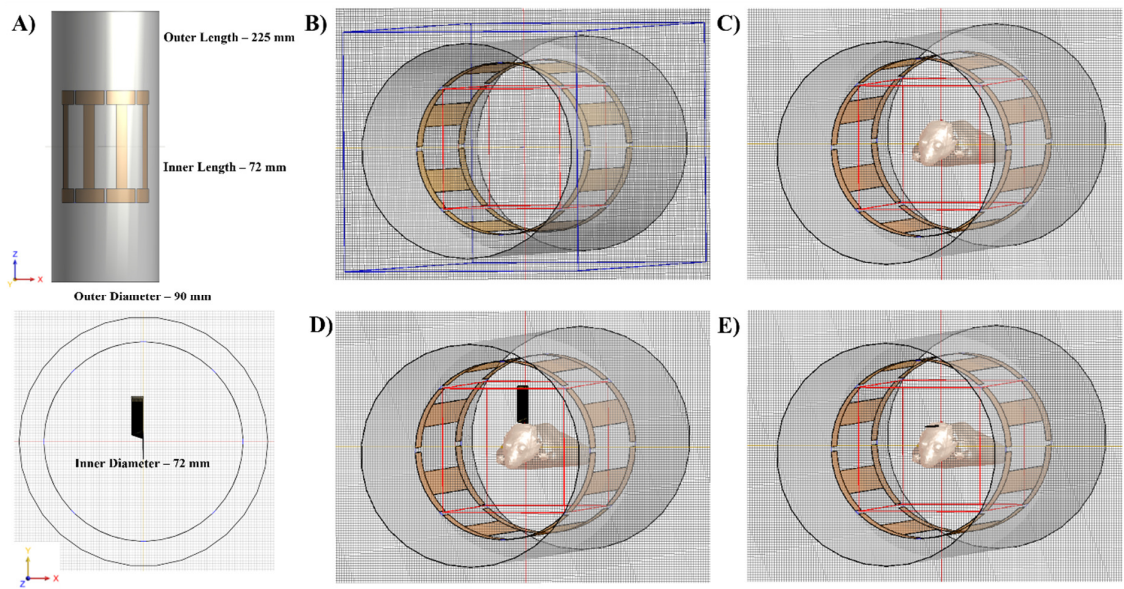

Figure 2. (A) Model of the 8-rung high-pass birdcage RF coil; Huygens' box approach: First step: creation of the Huygens' source, (B) with the RF coil acting as the source and the red box indicating the region of interest for the Huygens' box; Second step: Implementation of the Huygens' source with the mouse and probe models placed inside the Huygens' box; (C) Mouse model (without probe); Mouse model with (D) IC and (E) EC arrays.

## 2.2 Preclinical Models

The proposed HB method along with the MP reference simulations, was implemented in both mouse and rat anatomical models (Kainz *et al* 2006b), as shown in Fig. 1C–D.

*Mouse model*: B6C3F1 strain; body length: 80 mm; weight: 22.3 g; number of tissues: 68.

*Rat model*: Sprague Dawley strain; body length: 205 mm; weight: 385 g; number of tissues: 77.

These models were used to simulate the presence of IC and EC graphene arrays, enabling evaluation across different species and anatomical scales.

*2.3 EM Modelling and Simulation*

EM simulations were performed using *Sim4Life* (V8.0, ZMT, Switzerland), on a Windows 11 PC (3.00 GHz, 32 GB RAM, 23 GB NVIDIA 4090 GPU) using an excitation source parameter of 300 MHz Gaussian excitation with a bandwidth of 625 MHz in the presented and multi-port methods. The quantities estimated included RF transmit ($B_1^+$) field distribution, electric field ($E$), and SAR distribution.

A RF circuit co-simulation approach was implemented to estimate the optimal capacitor values for the transmission RF coil (Kumar *et al* 2021). This approach involves replacing all lumped elements with excitation ports and using full-port S-parameters generated via EM simulation to calculate the optimal resonating capacitor. This method prevents unwanted SAR variations caused by decoupling between excitation ports, which can affect transmit power efficiency and SAR deposition (Kumar *et al* 2021).

A cylindrical phantom with a diameter of 60 mm and a height of 80 mm, possessing electrical properties equivalent to brain tissue, including relative permeability ($\mu_r$) of 1, permittivity ($\varepsilon_r$) of 59.72, and electrical conductivity ($\sigma$) of 0.973235 S/m, was used as uniform loading for the calculation of the tuning capacitor of the RF coil. The simulated RF coil is of a two-port high-pass birdcage design, replicating a commercial coil (T11070V3, Bruker BioSpin MRI, Ettlingen, Germany) and consisting of eight rungs with a coil diameter of 72 mm and length of 72 mm, shield diameter of 90 mm and length of 225 mm, and rung width of 9.9 mm (Figure 2).

*Multi-port (MP) and HB methods:* The MP method comprises the entire system (RF coil, rodent model, and the probes if present), and the simulation is performed for each port of the RF coil individually (hence the term multi-port). Prior to post-EM analysis for MP, the *Sim4Life MATCH* toolbox was used to implement a matching circuit for fine tuning and impedance matching the RF coil to 50 ohms, and the fields are produced in circularly polarized combination. Sub-gridding is applied during the meshing step to obtain high-resolution meshes for the probes. HB simulations with the sub-gridding technique were conducted in two-steps. First, the HB source was generated by exciting the RF coil in circularly polarized mode with a 90° phase difference between the two ports (single-port excitation), producing fields in an empty rectangular region of interest (the HB). Second, this HB source was used to estimate the EM fields in configurations of interest, including the presence of IC or EC arrays. Sub-gridding was applied to achieve high-resolution meshes in the probe regions (Figure 2). Both MP and

HB simulations were conducted for mouse and rat models to evaluate the performance and accuracy of the HB method across different anatomical scales. Table I summarizes the simulation configurations, including the presence or absence of probes in both rodent models.

The EM fields were calculated, including $B_1^+$, $E$, and SAR. The total input power was normalized to produce a $B_1^+$ amplitude of 2μT at the center of the RF coil for each simulation method. Induced electric fields on probes were assessed, and mass-averaged SAR and peak spatially averaged SAR were calculated for 0.1g tissue, in accordance with the IEC guidelines (IEC 62704-1 2017).

Table 1. Comparison of Grid Size and Computational Runtime for Each Simulation Type in Mouse (A) and Rat (B) Models. MP simulations could not be completed at the same high-resolution grids as the HB approach on a single GPU. Only higher-resolution simulations are shown to highlight the computational advantages of the HB method.

### A) MOUSE

| | MP No probe | MP IC probe | HB Source Creation | HB No probe | HB IC probe | HB EC probe |
|---|---|---|---|---|---|---|
| Model View | 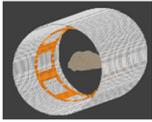 | 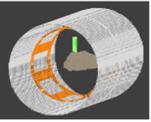 | 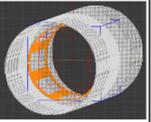 | 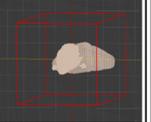 | 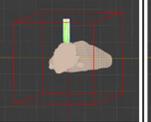 | 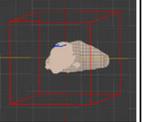 |
| Grid Size (MCells) | 82.5 | 49.5 | 17.5 | 28 | 110 | 31.3 |
| Voxel Size (mm³)<br>• gFET probe<br>• Rodent<br>• RF coil | [No probe]<br>0.25 x 0.25 x 0.25<br>1 x 1 x 1 | 0.01 x 0.01 x 0.0001<br>2 x 2 x 2<br>1 x 1 x 1 | [No probe]<br>1 x 1 x 1<br>1 x 1 x 1 | [No probe]<br>0.25 x 0.25 x 0.25<br>- | 0.01 x 0.01 x 0.0001<br>0.25 x 0.25 x 0.25<br>- | 0.0001 x 0.01 x 0.01<br>0.5 x 0.5 x 0.5<br>- |
| Run-time (hours) | ~6 | ~450 | ~0.5 | ~1 | ~220 | ~125 |

### B) RAT

| | MP No probe | MP IC probe | HB Source Creation | HB No probe | HB IC probe | HB EC probe |
|---|---|---|---|---|---|---|
| Model View | 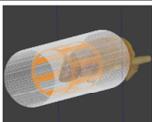 | 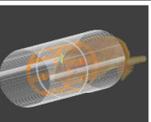 | 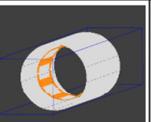 | 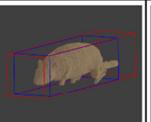 | 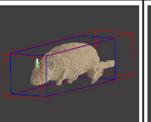 | 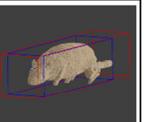 |
| Grid Size (MCells) | 52.5 | 54.3 | 30 | 35.6 | 76.5 | 63 |
| Voxel Size (mm³)<br>• gFET probe<br>• Rodent<br>• RF coil | [No probe]<br>0.5 x 0.5 x 0.5<br>1 x 1 x 1 | 0.01 x 0.01 x 0.0001<br>2 x 2 x 2<br>1 x 1 x 1 | [No probe]<br>1 x 1 x 1<br>1 x 1 x 1 | [No probe]<br>0.5 x 0.5 x 0.5<br>- | 0.01 x 0.01 x 0.0001<br>0.5 x 0.5 x 0.5<br>- | 0.0001 x 0.01 x 0.01<br>0.5 x 0.5 x 0.5<br>- |
| Run-time (hours) | ~4 | ~1050 (est.)<br>Unable to finish | ~1 | ~0.75 | ~203 | ~144 |

Note: A resolution of 1 × 1 × 1 mm³ for the HB is used during Huygens' source creation.

## 2.4 Uncertainty Analysis

To evaluate the reliability of the simulation methods, additional analyses were conducted to assess the sensitivity of $B_1^+$, $E$, and SAR outputs:

1. MP simulations with varying grid resolutions were performed.

2. HB simulations with different box lengths were used.

3. HB simulations under varying grid resolutions were conducted.

   A summary of the parameters assessed in the uncertainty analysis is provided in Table II.

Table 2. Summary of uncertainty analysis parameters. (A) MP simulations with varying grid resolutions. (B) HB simulations with different box lengths and voxel resolutions to assess the sensitivity and accuracy of $|B1^+|$, $|E|$, and SAR.

| A) | MP – No probe (evaluating the effect of voxel size) | | | | | |
|---|---|---|---|---|---|---|
| | Rodent voxel size (mm³) | 2 x 2 x 2 | 1 x 1 x 1 | 0.75 x 0.75 x 0.75 | 0.5 x 0.5 x 0.5 | 0.25 x 0.25 x 0.25 |

| B) | HB – No probe (evaluating the effect of Huygen's box size as well as the effect of voxel size) | | | | |
|---|---|---|---|---|---|
| | Box dimension (X × Y × Z mm³) | 50 x 50 x 100 | 100 x 100 x 250 | 100 x 100 x 300 | 100 x 100 x 350 |
| | Rodent voxel size (mm³) | • 1 x 1 x 1<br>• 0.75 x 0.75 x 0.75<br>• 0.5 x 0.5 x 0.5<br>• 0.25 x 0.25 x 0.25 | • 1 x 1 x 1<br>• 0.75 x 0.75 x 0.75<br>• 0.5 x 0.5 x 0.5<br>• 0.25 x 0.25 x 0.25 | • 1 x 1 x 1<br>• 0.75 x 0.75 x 0.75<br>• 0.5 x 0.5 x 0.5<br>• 0.25 x 0.25 x 0.25 | • 1 x 1 x 1<br>• 0.75 x 0.75 x 0.75<br>• 0.5 x 0.5 x 0.5<br>• 0.25 x 0.25 x 0.25 |

*2.5 Agreement and Statistical Analysis*

The performance and accuracy of the HB method were assessed by comparison with the full-wave multi-port simulations. The comparison focused on three EM quantities: $B_1^+$, $E$, and SAR. For each simulation case, including MP as reference, HB without implant, HB with IC, and HB with EC arrays, voxel-wise distributions of $B_1^+$, $E$, and SAR were extracted. The fields were mapped onto their respective voxel grids, preserving the native resolution of each simulation. Regions outside the anatomical model were masked to restrict the analysis to relevant tissue volumes.

Quantitative agreement between the HB and MP simulations was evaluated using Bland–Altman analysis. For each field metric, voxel-wise differences were plotted against the voxel-wise mean, enabling visualization of systematic deviations and dispersion. The following statistical measures were calculated for each case and field:

1. Bias (mean difference): representing the average voxel-wise deviation.

2. 95% Limits of Agreement (LoA): defined as the bias ± 1.96 times the standard deviation, indicating the range within which 95% of voxel differences are expected to lie.

In addition, percentage difference metrics (Δ%) were computed to provide relative comparison across simulations. The median Δ% and interquartile range (IQR) of Δ% were calculated for each field to summarize the central tendency and spread of voxel-wise differences. These metrics quantify both the magnitude and variability of deviations across cases and field types.

Bland–Altman plots illustrated systematic differences and the spread of voxel-wise deviations for $B_1^+$, $E$, and SAR. All post-processing and quantitative analyses were implemented in MATLAB (R2023b, MathWorks, Natick, MA, USA). For each simulation case, voxel-wise distributions of $B_1^+$, $E$, and SAR were extracted and processed to generate slices in axial, sagittal, and coronal orientations. To facilitate visual comparison, field distributions were displayed using absolute values for $B_1^+$ (T) and $E$ (V/m), and SAR (W/kg) in physical units. This combined statistical and visual analysis provided a robust framework for evaluating the accuracy and reliability of the HB method relative to full-wave MP simulations, with particular emphasis on the spatial and numerical consistency of $B_1^+$, $E$, and SAR distributions in the presence of IC and EC arrays.

## 3. Results

### 3.1 EM Field Distribution and Quabntitative Metrics

The computational demands differed markedly between the simulation approaches. HB simulations required 125 to 220 hours, compared to ~450 to >1050 hours for MP simulations across different grid sizes (Table I). Summary metrics of EM field maps and safety-relevant quantities are reported in Table III for both mouse and rat models. The comparison includes mean $|B_1^+|$ peak electric field magnitude ($|E_{max}|$), mass-averaged SAR, and peak spatially averaged SAR over 0.1 g of tissue. MP simulations with IC arrays were performed only for the mouse model; MP simulations for the mouse model with EC arrays or rat model with IC or EC arrays were not completed due to computational limits.

Figures 3 and 5 show spatial distributions $|B_1^+|$, $|E|$, and SAR for the mouse and rat models, respectively, across central axial, coronal, and sagittal slices. Quantitative evaluation confirmed strong agreement between HB and MP simulations: mean $|B_1^+|$ values differed by less than 2%, maximum electric field deviations reached approximately 8% without the probe and 11% with the probe, and $SAR_{max}$ differences were ~0.02–0.05 W/kg. Importantly, all SAR values remained well below human safety limits (3.2 W/kg whole-head, 10 W/kg local), indicating that the observed deviations are unlikely to pose safety concerns in rodents.

### 3.2 Uncertainty Analysis

Varying the grid resolution between 0.25 mm and 1 mm in the MP simulations affected RF port decoupling (or isolation), which ranged from -26 to -46 dB when the tuning capacitor values were kept the same. This variation led to a 37% and 41% decrease in $SAR_{avg}$ and $SAR_{max}$, respectively, when using a 0.25 mm resolution. For the HB method, varying the box size revealed that inaccurate EM field generation occurred at the box boundary when its size was smaller than the overall RF coil model dimensions, with no significant changes observed when grid resolutions varied between 0.25 and 1 mm.

Table 3. Summary of $|B_1^+|_{mean}$, peak electric field magnitude ($|E|_{max}$), mass-averaged SAR ($SAR_{avg}$), and peak spatially averaged SAR ($SAR_{max}$) over 0.1 g of tissue in the head region across different simulation configurations for (A) mouse and (B) rat models.

A)

| MOUSE | $|B_1^+|$ Mean (µT) | $|E|$ Max (V/m) | Mass-avg SAR (watt/kg) | Peak spatial-avg SAR (watt/kg) |
|---|---|---|---|---|
| MP without array | 1.930 | 69.7 | 0.081 | 0.18 |
| MP with IC | 1.928 | 1968.6 | 0.088 | 0.19 |
| HB without array | 1.929 | 75.5 | 0.085 | 0.21 |
| HB with IC | 1.929 | 1956.3 | 0.090 | 0.24 |
| HB with EC | 1.929 | 2070.4 | 0.092 | 0.234 |

B)

| RAT | $|B_1^+|$ Mean (µT) | $|E|$ Max (V/m) | Mass-avg SAR (watt/kg) | Peak spatial-avg SAR (watt/kg) |
|---|---|---|---|---|
| MP without array | 1.80 | 208.4 | 0.074 | 4.28 |
| HB without array | 1.77 | 350.08 | 0.102 | 4.32 |
| HB with IC | 1.82 | 6970.3 | 0.109 | 5.27 |
| HB with EC | 1.83 | 4911.9 | 0.114 | 5.46 |

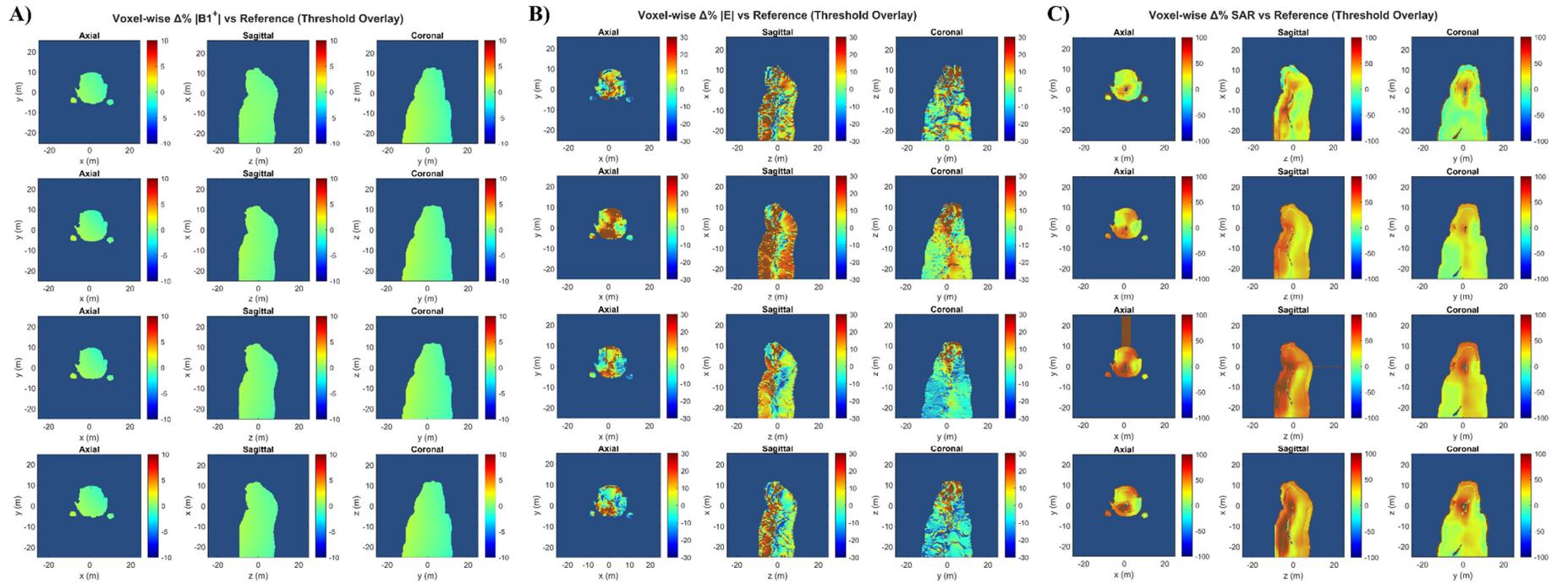

Figure 3. Simulated $|B_1^+|$, $|E|$ and SAR field distributions in the mouse model for MP and HB configurations. MP simulations are shown without and with IC arrays, while HB simulations are shown without arrays, with IC arrays, and with EC arrays. Field maps are presented in central (A) axial, (B) coronal, and (C) sagittal slices. All values are normalized to the input power required to generate a $|B_1^+|$ amplitude of 2 µT at the center of the RF coil.

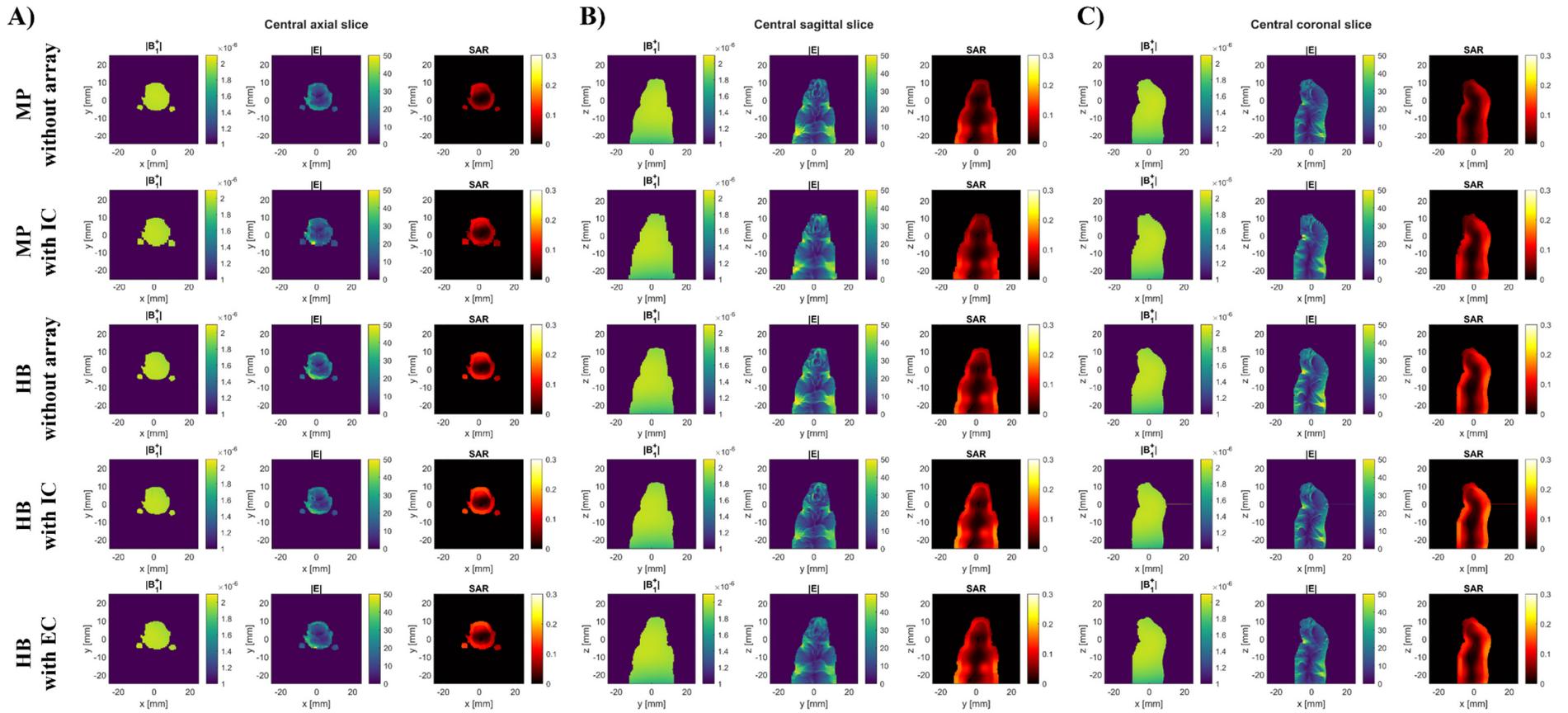

Figure 4. Bland–Altman analysis of $|B_1^+|$, $|E|$ and SAR fields in the mouse model. MP simulations with IC arrays and HB simulations without arrays, with IC arrays, and with EC arrays are compared, using MP without arrays as the reference. Field maps are shown in central (A) axial, (B) coronal, and (C) sagittal slices.

## 3.3 Agreement and Statistical Analysis

Figures 4 and 6 show Bland–Altman plots and agreement metrics derived from Bland–Altman analysis, including bias, 95% LoA, median Δ%, and IQR of Δ%, were tabulated for each field and simulation case, as shown in Table IV-V. These tables provide a concise numerical summary of the agreement between the HB and MP simulations.

Table 4. Bland–Altman and distribution-based agreement metrics for mouse model simulations. Metrics include bias, 95% limits of agreement (LoA), median percentage difference (Δ%), and interquartile range (IQR) of Δ%, comparing MP and HB configurations across different simulations.

**A)** $|B_1^+|$ (T)

| Case | Bias (mean diff) | LoA (lower–upper) | Median Δ% | IQR Δ% |
|---|---|---|---|---|
| MP with IC | -1.88e-09 | -3.60e-08 – 3.22e-08 | -0.06 | 1.29 |
| HB without array | -5.86e-10 | -3.46e-08 – 3.35e-08 | 0.07 | 1.42 |
| HB with IC | -1.07e-09 | -3.50e-08 – 3.29e-08 | 0.05 | 1.42 |
| HB with EC | -1.33e-09 | -3.61e-08 – 3.35e-08 | 0.03 | 1.38 |

**B)** $|E|$ (V/m)

| Case | Bias (mean diff) | LoA (lower–upper) | Median Δ% | IQR Δ% |
|---|---|---|---|---|
| MP with IC | 0.78 | -9.44 – 10.99 | 6.24 | 25.83 |
| HB without array | 2.65 | -3.99 – 9.29 | 12.89 | 23.09 |
| HB with IC | -0.25 | -7.55 – 7.05 | -1.36 | 20.38 |
| HB with EC | -0.44 | -8.44 – 7.55 | -1.71 | 23.20 |

**C)** SAR (W/kg)

| Case | Bias (mean diff) | LoA (lower–upper) | Median Δ% | IQR Δ% |
|---|---|---|---|---|
| MP with IC | 0.015 | -0.04 – 0.07 | 16.23 | 31.46 |
| HB without array | 0.02 | -0.02 – 0.07 | 31.74 | 38.98 |
| HB with IC | 0.03 | -0.02 – 0.08 | 48.41 | 43.71 |
| HB with EC | 0.03 | -0.03 – 0.09 | 43.10 | 44.14 |

***Mouse model:***

- $|B_1^+|$: Biases were negligible (~$10^{-9}$ T) with narrow LoA. Median Δ% was 0.02–0.07%,

IQR ~1.3–1.4%.

- $|E|$: Median Δ% ranged from −1.7% to +13%, IQR ~20–25%.

- *SAR*: Median Δ% ranged ~16% to ~48%, IQR up to ~44%. Absolute differences were ~0.02–0.03 W/kg.

*Rat model:*

- $|B_1^+|$: Biases were negligible (~$10^{-8}$ T) with narrow limits of agreement. Median Δ% was −3.7% to −1.7%, IQR ~4.6–5.2%.

- $|E|$: Median Δ% ranged from ~5.1% to 8.4%, IQR ~20–23%.

- *SAR*: Median Δ% ranged ~7.0% to 12.5%, IQR up to ~33%. Absolute differences were ~0.03–0.05 W/kg.

Table 5. Bland–Altman and distribution-based agreement metrics for rat model simulations. Metrics include bias, 95% limits of agreement (LoA), median percentage difference (Δ%), and interquartile range (IQR) of Δ%, comparing MP and HB configurations across different simulation setups.

**A)** $|B_1^+|$ (T)

| Case | Bias (mean diff) | LoA (lower–upper) | Median Δ% | IQR Δ% |
|---|---|---|---|---|
| HB without array | -8.85e-08 | -3.15e-07 – 1.38e-07 | -3.70 | 4.65 |
| HB with IC | -5.20e-08 | -2.83e-07 – 1.79e-07 | -1.71 | 4.79 |
| HB with EC | -5.62e-08 | -2.80e-07 – 1.68e-07 | -1.93 | 5.17 |

**B)** |E| (V/m)

| Case | Bias (mean diff) | LoA (lower–upper) | Median Δ% | IQR Δ% |
|---|---|---|---|---|
| HB without array | 1.72 | -12.26 – 15.69 | 5.11 | 20.28 |
| HB with IC | 2.61 | -11.59 – 16.82 | 8.44 | 20.11 |
| HB with EC | 1.97 | -11.77 – 15.72 | 7.06 | 22.66 |

**C)** SAR (W/kg)

| Case | Bias (mean diff) | LoA (lower–upper) | Median Δ% | IQR Δ% |
|---|---|---|---|---|
| HB without array | 0.03 | -0.14 – 0.19 | 6.99 | 29.64 |
| HB with IC | 0.05 | -0.13 – 0.23 | 12.45 | 30.44 |
| HB with EC | 0.03 | -0.15 – 0.22 | 9.03 | 33.18 |

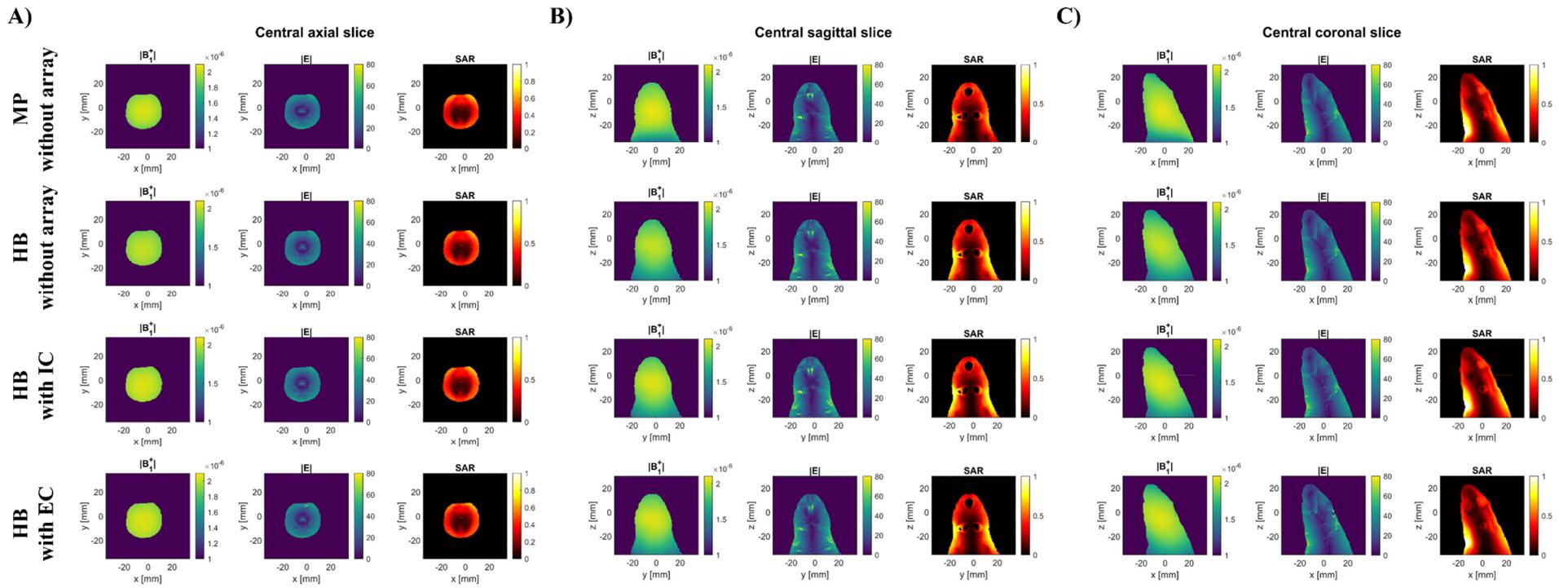

Figure 5. Simulated |B1+|, |E| and SAR field distributions in the rat model for MP and HB configurations. MP simulations are shown without arrays, while HB simulations are shown without arrays, with IC arrays, and with EC arrays. Field maps are presented in central (A) axial, (B) coronal, and (C) sagittal slices. All values are normalized to the input power required to generate a |B1+| amplitude of 2 µT at the center of the RF coil.

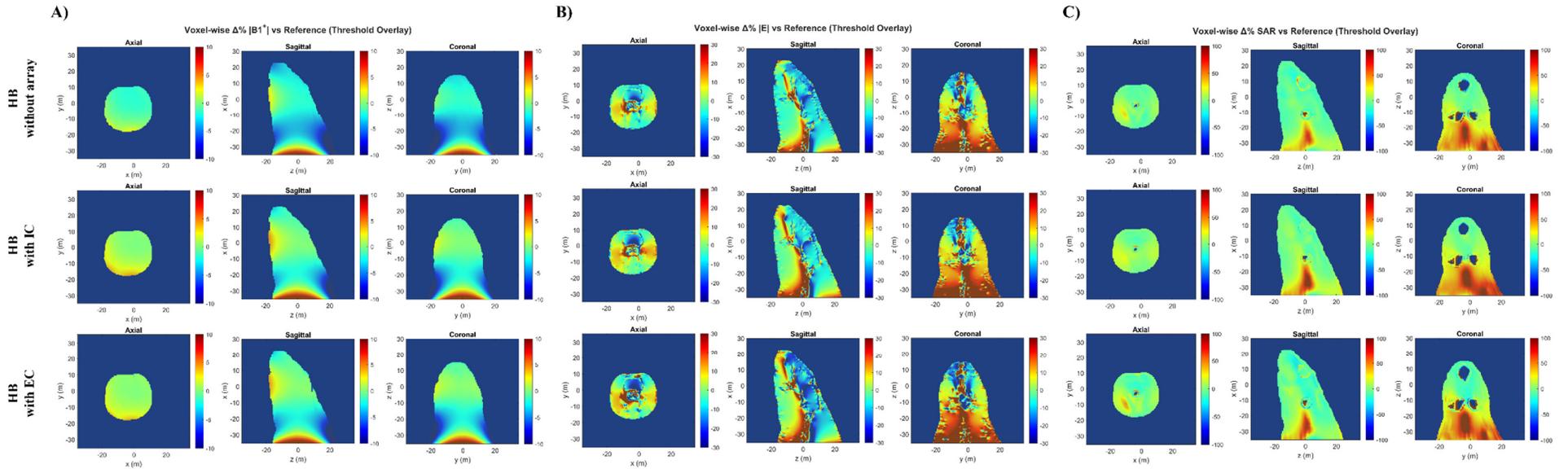

Figure 6. Bland–Altman analysis of $|B_1^+|$, $|E|$ and SAR fields in the rat model. HB simulations without arrays, with IC arrays, and with EC arrays are compared, using MP without arrays as the reference. Field maps are shown in central (A) axial, (B) coronal, and (C) sagittal slices.

## 4. Discussion

*4.1 Overview and Motivation*

To address the challenge of impractical simulation times and sub-optimal accuracy in systems containing microscopic array structures several orders of magnitude smaller than the RF coil, this work implemented and evaluated an accelerated hybrid EM simulation methodology based on the HB approach with sub-gridding. The method was applied to evaluate EM field distributions in animal models, with and without implanted microscopic arrays. By replacing direct full-wave MP excitation with equivalent sources on a closed surface, HB minimizes the computational domain while maintaining high-resolution accuracy, offering a feasible alternative for complex implant and RF coil interactions. Although HB does not reproduce MP results exactly, it preserves overall field distributions and spatial patterns critical for evaluating RF performance and safety in MRI systems with implants.

*4.2 Computational Performance and Field Accuracy*

Our results demonstrate that HB simulations substantially reduced computational runtime by approximately 70-80%. For example, HB simulations were completed in under 220 hours even with very fine mesh resolutions, whereas MP simulations required approximately 450 hours with IC arrays with very coarse resolution and could not be performed for EC arrays or for the rat model due to excessive computational demands for a single GPU. This computational efficiency is particularly important when modelling high-density arrays, where it is challenging or impractical with conventional simulation approach. Despite this acceleration, HB maintained $|B_1^+|$ field symmetry and preserved localized $|E|$ and SAR patterns comparable to MP results. High-resolution field estimation is achieved without excessive computational cost, while preserving essential spatial characteristics of $|B_1^+|$, $|E|$, and SAR.

The performance and accuracy of HB were quantitatively evaluated using Bland–Altman analyses against MP simulations. Unlike approaches relying solely on CoV to assess $|B_1^+|$ field uniformity, Bland–Altman provides voxel-wise comparison, allowing assessment of both biases and local deviations. This method quantifies the mean difference and limits of agreement, offering statistical confidence in HB predictions and highlighting spatial regions where discrepancies may occur, such as near coil edges or tissue boundaries. For context, CoV values for $|B_1^+|$ were 3.6–3.7% in the mouse model and 8.9–9.4% in the rat model, reflecting global field variability (expressing CoV as a percentage also implies proportionality between variability and the mean of measurements (Martin Bland and Altman 1986)). The Bland–

Altman analyses confirmed small biases and narrow limits of agreement for $|B_1^+|$, with somewhat broader distributions for |E| and SAR due to their higher spatial variability.

Across all configurations, $|B_1^+|$ distributions were highly consistent in both mouse and rat models, with 0–1.5% of voxels exceeding a 10% deviation threshold. |E| exhibited minor localized deviations, with 3.9–4.6% of voxels in the mouse and 6–7.5% in the rat exceeding 10%, mainly near tissue or probe interfaces. SAR deviations remained low, with less than 6.3% of voxels exceeding 20% in the mouse and below 5% in the rat, indicating that small local E-field variations were amplified but remained within acceptable limits. These results demonstrate that HB reproduces MP field characteristics with high fidelity, validating its reliability for safety-relevant field analysis while offering a significant computational advantage.

*4.3 Graphene-Based Arrays and Safety Implications*

HB simulations were used to evaluate SAR deposition from graphene-based arrays in the MRI environment. The arrays exhibited only a modest increase in SAR. The hybrid HB sub-gridding approach significantly improves simulation efficiency while maintaining close agreement with MP results, particularly around the arrays and within the rodent brain. Although standardized SAR limits for rodents are not established, human safety guidelines specify a maximum whole-head SAR of 3.2 W/kg and a local SAR of 10 W/kg (IEC/FDA). All SAR values observed in this study were well below these thresholds, indicating that the deviations are unlikely to pose safety concerns.

The safety assessment performed here is essential to ensure that elevated SAR or thermal doses do not compromise brain function. RF exposure has been associated with changes in neuronal excitability, dendritic spine morphology, and recognition memory (Liu *et al* 2024, Narayanan *et al* 2019, Wang *et al* 2017). Such changes could compromise the accuracy of brain recordings and confound disease-related findings in both healthy and diseased rodent models (Eskandani and Zibaii 2023, Abtin *et al* 2024). The modest SAR elevation observed in this work therefore supports the suitability of graphene arrays for preclinical multimodal MRI–electrophysiology research.

*4.4 Scalability and Application to Complex Models*

Cross-model comparisons further illustrate HB's scalability. $|B_1^+|$ distributions showed minimal deviations relative to MP across both mouse and rat models, while |E| and SAR fields exhibited local variations influenced by array placement and tissue heterogeneity. Greater

variability in the rat model reflects larger anatomical dimensions and more complex probe configurations, demonstrating HB's ability to handle increased model complexity.

In the rat model, MP simulations with IC arrays were computationally prohibitive, whereas HB simulations remained feasible. This makes HB a practical alternative for studies where MP modelling is unfeasible, enabling systematic parameter sweeps and high-resolution analyses across multiple rodent models, with potential scalability to larger animal and even human models.

*4.5 Limitations and Future Directions*

While promising, the HB approach has several limitations. First, RF coil matching is currently not possible in HB as it is applicable in MP, introducing slight uncertainties. Second, uncertainty can arise from Huygens' box size, which must be sufficiently large to avoid cross-talk between the source and the model. Third, sub-gridding implementation requires careful interface handling to avoid discontinuities or numerical instability (Hartley *et al* 2022).

Another limitation concerns the electrical properties of graphene, the dielectric and conductive properties of single layer graphene used here were based on reported values from the American Chemical Society (Anon 2025), providing a reasonable baseline for simulation. However, these properties can vary significantly depending on structural and environmental factors such as temperature, chemical potential (Fermi level), relaxation time, doping level, and layer thickness (Sreeprasad and Berry 2013, Lalire *et al* 2024). Conductivity is one of the most important parameters governing graphene's electrical applications (Chen *et al* 2023), and can be engineered for specific applications. Future studies could integrate experimentally derived parameters, example from electrical impedance tomography (Khambampati *et al* 2020), further refining the simulations.

*4.6 Outlook and Significance*

Graphene-based arrays are emerging as promising tools for multimodal neuroimaging and electrophysiology. However, their small dimensions and complex geometries challenge conventional EM modeling approaches. The proposed work addresses these challenges and offers an efficient method for evaluating microscopic implants in multimodal MRI setting. By maintaining key spatial characteristics of $|B_1^+|$, $|E|$, and SAR distributions while reducing computational cost by up to 80%, HB enables detailed preclinical safety studies that are otherwise impractical with full-wave MP simulations. Furthermore, the methodology proposed here can be adapted to other microscopic probes/arrays and extended to future translational and clinical research, advancing both safety evaluation and implant design in the MRI environment.

## 5. Conclusion

This study demonstrates the feasibility and effectiveness of the HB method for evaluating electromagnetic interactions between graphene-based electrophysiology arrays and MRI RF coils. By employing equivalent sources on a closed surface combined with sub-gridding, the HB approach substantially reduces computational requirements, enabling simulations that would otherwise be prohibitive with full-wave multi-port modeling, particularly in the presence of implanted arrays. The method preserves key spatial features of the $|B_1^+|$, $|E|$, and SAR distributions, providing reliable RF and SAR estimates across rodent models while maintaining high computational efficiency. The approach is scalable and adaptable to various probe geometries and material parameters, making it a powerful framework for preclinical MRI safety evaluation. Moreover, its computational efficiency facilitates the accelerated translation of microscopic implant simulations to human-scale modeling, supporting future developments in multimodal neuroimaging and implant design.

**Data availability**

Available from the corresponding author upon reasonable request.

**Acknowledgements**

This work was supported in part by the UK Engineering and Physical Sciences Research Council (EPSRC) under Grant EP/X013669/1. The authors gratefully acknowledge Sim4Life, ZMT, for providing an academic license.